\documentclass[aps,12pt,preprintnumbers,nofootinbib,superscriptaddress]{revtex4}
\usepackage{graphicx}
\usepackage{amssymb,amsmath}
\usepackage{slashed}
\usepackage{dcolumn}
\def\Dslash{D\hskip-0.65em /}

\begin{document}

\title{\Large{\bf Electrodynamics Modified by Some Dimension-five Lorentz Violating Interactions: Radiative Corrections } }

\author{Shan-quan Lan}
\author{Feng Wu}
\email[Electronic address: ]{fengwu@ncu.edu.cn}
\affiliation{%
Department of Physics, Nanchang University,
330031, China}

\date{\today}

\begin{abstract}
We study radiative corrections to massless quantum electrodynamics modified by two dimension-five LV interactions $\bar{\Psi} \gamma^{\mu} b'^{\nu} F_{\mu\nu}\Psi$ and $\bar{\Psi}\gamma^{\mu}b^{\nu} \tilde{F}_{\mu\nu} \Psi$ in the framework of effective field theories. All divergent one-particle-irreducible Feynman diagrams are calculated at one-loop order and several related issues are discussed. It is found that massless quantum electrodynamics modified by the interaction $\bar{\Psi} \gamma^{\mu} b'^{\nu} F_{\mu\nu}\Psi$ alone is one-loop renormalizable and the result can be understood on the grounds of symmetry. In this context the one-loop Lorentz-violating beta function is derived and the corresponding running coefficients are obtained.
\end{abstract}
\pacs{} 
\maketitle 
\newpage

\date{\today}

\section{Introduction}
Despite its success to account for nearly all phenomena with precision down to subatomic scales, the standard model of particle physics is incomplete and leaves several issues unsolved. Beyond the standard model, exploring the possible new physics involving a violation of Lorentz symmetry is an interesting and extensively studied subject in recent years. In particular, Colladay and Kosteleck$\acute{\rm y}$  \cite{Colladay} have systematically constructed Lorentz-violating (LV) terms of renormalizable dimensions and many related contents have been intensely investigated \cite{CFJ, Coleman, fermion, radiative, loop, CPT1, ODD, EVEN}.

Nevertheless, the fact that no significant departure from Lorentz invariance has been observed in precision tests raises a subtle ``Lorentz fine-tuning problem'' \cite{FT} in this context. One possible resolution is that the currently unknown underlying theory prohibits the generation of the renormalizable LV operators at low energies. Probing this scenario at high-energy scales would be interesting but lies beyond the scope of this paper. However, this conception raises the interest in studying the LV terms of nonrenormalizable dimensions.

Studies in the literature of nonrenormalizable LV operators are relatively scanty. In the framework of effective field theories, a nonrenormalizable theory treated as a low energy effective field theory, valid up to some mass scale $M$ of new physics, might still be sensible and reliable predictions could be made from it. At low energies, effects due to nonrenormalizable terms are suppressed by inverse powers of $M$. This power suppression makes nonrenormalizable operators ``safer" than the renormalizable ones. A few investigations have been carried out in this direction \cite{HD1, HD2, HD3, Gomes, Lan}.  

In this work, we focus on quantum electrodynamics (QED) modified by two dimension-five LV interactions $\bar{\Psi} \gamma^{\mu} b^{\nu} F_{\mu\nu}\Psi$ and $\bar{\Psi}\gamma^{\mu}b^{\nu} \tilde{F}_{\mu\nu} \Psi$, where $\tilde{F}_{\mu\nu} \equiv {1\over 2} \epsilon_{\mu\nu\alpha\beta} F^{\alpha\beta}$ is the dual electromagnetic tensor and the fixed vector background $b^{\mu}$ is assumed to be the only source that induces the Lorentz symmetry breaking. Several issues related to these two LV terms have been studied and non-trivial results are obtained \cite{HD2, HD3, Gomes, Lan}. In particular, it is found that with the operator $ \bar{\Psi}\gamma^{\mu}b^{\nu} \tilde{F}_{\mu\nu} \Psi $, a charged spinor possesses a spin-independent magnetic dipole moment density, along with the usual one associated with its spin. Also, the degeneracy of the hydrogen energy spectrum is shown to be completely removed by the $CP$-even operator $\bar{\Psi}\gamma^{i}b^{j} \tilde{F}_{ij} \Psi$. The LV operator $ \bar{\Psi} \gamma^{\mu} b^{\nu} F_{\mu\nu}\Psi $ takes no part in determining the atomic energy spectrum. For more details, see Ref.\cite{Lan}. 

From the field-theoretic point of view, it is interesting to study the quantum corrections of an effective theory containing nonrenormalizable LV terms. A general feature of a nonrenormalizable theory is that one would not be able to reabsorb all the ultraviolet (UV) divergent quantum corrections into the coupling constants in the original Lagrangian, and new counterterms permitted by symmetry are needed at each order of perturbative calculations. So far, even in the simplified case where massless QED is modified by the two non-minimal LV operators mentioned above, a comprehensive study of one-loop radiative corrections to this model is still lacking. Some one-loop calculations of the photon self-energy amplitude have been performed \footnote{It is claimed in \cite{Gomes} that an aetherlike term is radiatively generated by the operator $ \bar{\Psi}\gamma^{\mu}b^{\nu} \tilde{F}_{\mu\nu} \Psi $. However, the calculations leading to Eq. (65) in that paper is erroneous, and in fact no aetherlike term is generated in the case of electrodynamics. } \cite{Gomes}. The goal of this work is to fill this gap by determining all the divergent one-loop corrections and identify the higher dimensional counterterms that should be added to the Lagrangian at the beginning so that the theory is consistent at one-loop order. 

The rest of the paper is organized into three parts. In Sec. II, using the Feynman rules for massless QED modified by two non-minimal LV interactions $\bar{\Psi} \gamma^{\mu} b'^{\nu} F_{\mu\nu}\Psi$ and $\bar{\Psi}\gamma^{\mu}b^{\nu} \tilde{F}_{\mu\nu} \Psi$, the superficial degree of freedom of a general Feynman diagram is determined. We then compute all divergent radiative corrections to the Lagrangian at one-loop order and find out all new counterterms required in order to render the corrections finite. Some related issues are discussed along the way. In Sec. III, based on the results of Sec. II, we investigate the special case where massless QED is modified by only one LV operator $\bar{\Psi} \gamma^{\mu} b'^{\nu} F_{\mu\nu}\Psi$ and argue the renormalizability of the theory in this context. The one-loop beta function for the LV coefficients $b'^{\,\alpha}$ is derived and then used to solve for the running LV coefficients. Our conclusions are given in the final section.

\section{One-loop corrections}
We start with the LV model defined by the Lagrangian density as follows:
\begin{equation}
\mathcal{L}=-\frac{1}{4}F_{\mu\nu}F^{\mu\nu}+\bar{\Psi}(i\Dslash-\gamma^{\mu}b'^{\nu}F_{\mu\nu}-\gamma^{\mu}b^{\nu}\tilde{F}_{\mu\nu})\Psi,  \label{L} 
\end{equation}
where the gauge covariant derivative takes the form $D^{\mu} =\partial^{\mu} +i e A^{\mu}$ with $e$ being the gauge coupling constant determining the strength of the electromagnetic interaction. The mass dimension of the fixed vector background $b^{\mu}$ and $b'^{\mu}$ is $-1$. Here $b^{\mu}$ and $b'^{\mu}$ are chosen differently since we have absorbed possible coupling constants of the two LV terms into the fixed vector background.
 
Following the standard procedure, perturbative analysis begins with gauge fixing. Feynman rules for the fermion and photon propagators are the usual ones. With the introduction of the LV terms in the Lagrangian~(\ref{L}), the fermion-photon vertex is given by 
 \begin{equation}
V^{\mu}(q) = -ie\gamma^{\mu}+ b'\cdot{q}\gamma^{\mu}-b'^{\mu}\slashed{q}-\epsilon^{\mu}_{\,\,\alpha\beta{\nu}}b^{\alpha}\gamma^{\beta}q^{\nu}
\end{equation}
where $q^{\mu}$ is the photon momentum pointing into the vertex.

By naive power counting, the superficial degree of divergence $D$ of a Feynman diagram is
 \begin{equation}
D= 4-N_{\gamma} -{3\over 2} N_{e} +V, \label{power}
\end{equation}
where $N_\gamma$ is the number of external photon legs, $N_{e}$ is the number of external fermion legs, and $V$ is the number of fermion-photon vertices. At one-loop order, we only need to consider the part of the diagram that is either zeroth or first order in coefficients of Lorentz violation. It would be inconsistent to include terms that are nonlinear in LV coefficients without also considering multiloop diagrams which could contribute at the same order \cite{loop}. Hereafter, whenever we refer to one-loop, we mean the part of one-loop that is at most linear in LV coefficients.  

Notice that although the Lagrangian~(\ref{L}) violates $CPT$, it preserves $C$ parity. Therefore, the conventional Furry theorem holds and any vacuum expectation value of an odd number of currents vanishes. The one-loop four-point photon amplitude, in spite of having positive superficial degree of divergence from~(\ref{power}), is finite because of the requirement of gauge invariance.   
 
In summary, at one-loop order, there are four divergent one-particle-irreducible amplitudes, as shown in Fig. 1. For the remainder of this section, we will calculate all the divergent one-loop corrections. In order to extract the UV singularities, we adopt dimensional regularization to evaluate the integrals in $d=4-\epsilon$ dimensional spacetime. The applicability of standard dimensional regularization techniques in LV theories is discussed in \cite{loop}.   
\begin{figure}[t]
\begin{center}
\includegraphics[width=15cm,clip=true,keepaspectratio=true]{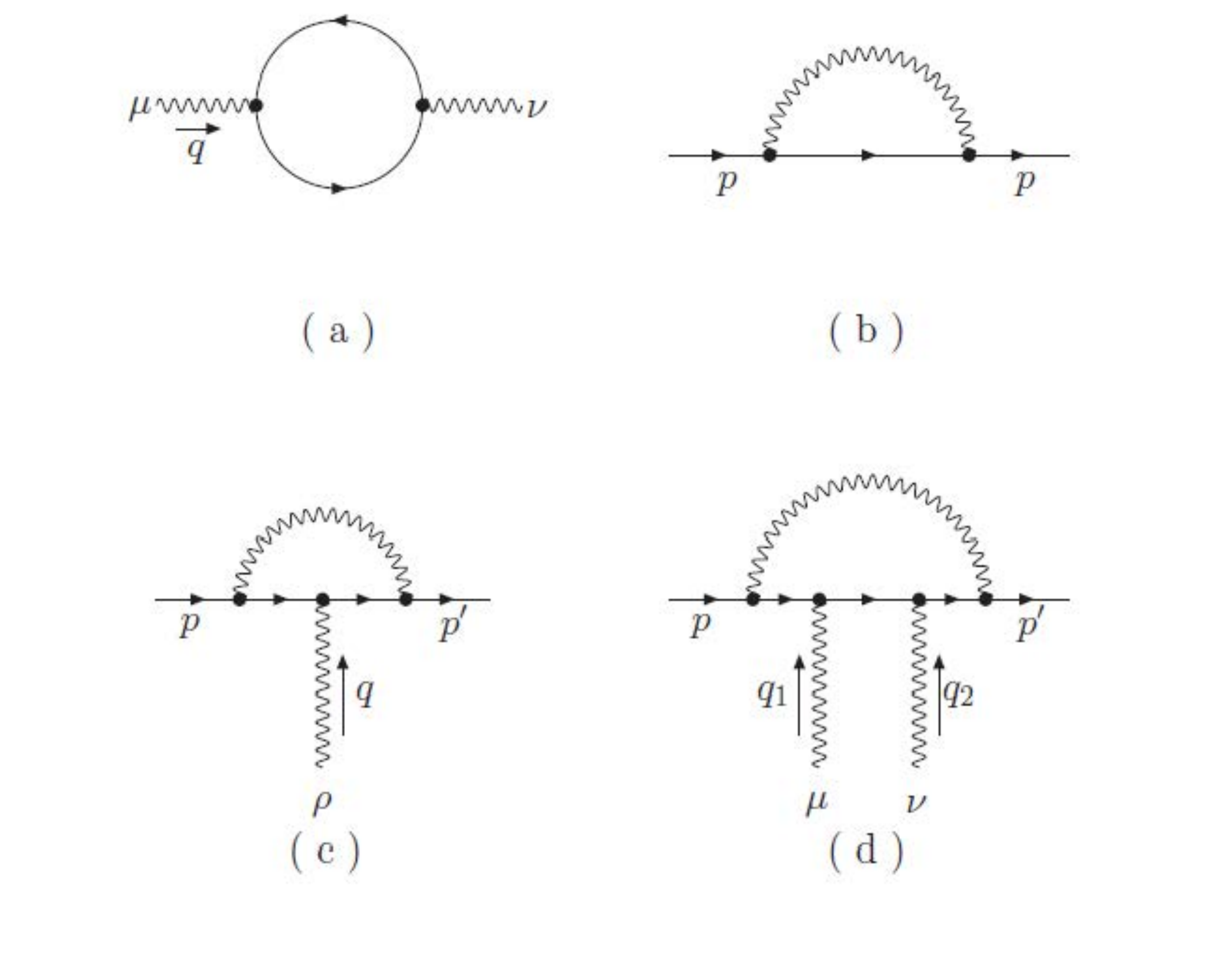}
\caption{\small The four one-loop amplitudes with UV divergences.}
\end{center}\label{loop}
\end{figure}
 
\subsection{Photon self-energy}
Applying the Feynman rules, an expression corresponding to the one-loop photon self-energy $i \Pi^{\mu\nu}(q)$ (Fig. 1a) is
 \begin{equation}
i \Pi^{\mu\nu}(q)=(-1) \int {d^d k \over (2\pi)^{d}} \rm{tr } \left( V^{\mu}(q) {i\over \slashed{k}} V^{\nu}(-q) {i\over \slashed{k}+\slashed{q}}\right).
\end{equation}  
After manipulating the Dirac matrices, we have
\begin{eqnarray}
i \Pi^{\mu\nu}(q)
&=&-ed\int\frac{\textrm{d}^{d}k}{(2\pi)^{d}}\frac{1}{k^{2}(k+q)^{2}}\nonumber\\
&\,\,&\{e[k^{\mu}(k+q)^{\nu}-g^{\mu\nu}k\cdot(k+q)+k^{\nu}(k+q)^{\mu}]\nonumber\\
&\,\,&+i\epsilon^{\nu}_{\,\,\alpha \beta \sigma}b^{\alpha}q^{\sigma}[k^{\mu}(k+q)^{\beta}-g^{\mu{\beta}}k\cdot(k+q)+k^{\beta}(k+q)^{\mu}]\nonumber\\
&\,\,&-i\epsilon^{\mu}_{\,\,\alpha\beta\rho}b^{\alpha}q^{\rho}[k^{\beta}(k+q)^{\nu}-g^{\beta\nu}k\cdot(k+q)+k^{\nu}(k+q)^{\beta}]\}.
\end{eqnarray}
Then, by means of standard steps including introducing a Feynman parameter, shifting the integration variable, and performing the momentum integral, we obtain
\begin{eqnarray}
i \Pi^{\mu\nu}(q)&=& {ie\over (4\pi)^{d/2}} 2d {\Gamma(2-{d\over 2})(\Gamma({d\over 2}))^2 \over \Gamma(d)} 
(-q^2)^{{d\over 2}-2} [ e (q^{\mu}q^{\nu} - g^{\mu\nu} q^2) -2ig^{\nu\alpha}g^{\mu\beta}\epsilon_{\alpha\beta\sigma\rho}b^{\rho}q^{\sigma}q^{2}]\nonumber\\
&=& {ie^2 \over 6\pi^2\epsilon}(q^{\mu}q^{\nu} - g^{\mu\nu} q^2)-{ie\over 3\pi^2\epsilon}g^{\nu\alpha}g^{\mu\beta}\epsilon_{\alpha\beta\sigma\rho}b^{\rho}q^{\sigma}q^{2}+\rm{ finite\,\, part}.\label{photon}
\end{eqnarray}

Some comments regarding this result are in order. First of all, current conservation guarantees that the result~(\ref{photon}) obeys the Ward-Takahashi identity. This can be seen by dotting the photon momentum $q^{\mu}$ into the amplitude~(\ref{photon}), which gives zero. Second, while the divergent term proportional to $i(q^{\mu}q^{\nu} - g^{\mu\nu} q^2) $ can be renormalized by the usual QED counterterm proportional to $F^{\mu\nu} F_{\mu\nu}$, the second term in~(\ref{photon}) shows that the one-loop correction to the photon-photon correlation function due to the LV operator $ \bar{\Psi}\gamma^{\mu}b^{\nu} \tilde{F}_{\mu\nu} \Psi $ generates a new type of divergence which cannot be absorbed in the original Lagrangian~(\ref{L}). It is straightforward to show that the new counterterm needed in order to cancel this divergence is of the form $b^{\alpha}F^{\mu\nu} \partial_{\mu} \tilde{F}_{\alpha\nu}$. This is the leading higher derivative term allowed by symmetries. Finally, our result shows that the LV operator $  \bar{\Psi}\gamma^{\mu}b'^{\nu} F_{\mu\nu} \Psi$ does not contribute to photon self-energy at this order.
 \subsection{Fermion self-energy}
 The one-loop diagram contributing to the fermion self-energy is shown in Fig. 1b. This contribution, denoted by $-i \Sigma (p) $, in Feynman gauge is given by
\begin{equation}
-i \Sigma ( p) = \int {d^d k \over (2\pi)^{d}}  V^{\mu}(p-k) {i \slashed{k}\over k^2} V^{\nu}(k-p) {-ig_{\mu\nu}\over (p-k)^2}.
\end{equation}  
By direct evaluation, we have
\begin{eqnarray}
-i\Sigma (p ) &=&-\frac{e}{(4\pi)^{\frac{d}{2}}}\int_{0}^{1}\textrm{d}x\{\Gamma(2-\frac{d}{2})\left(x(x-1)p^2\right)^{\frac{d}{2}-2}[ie(2-d)x\slashed{p}+2(1-x)x\slashed{p}\epsilon_{\mu\alpha\beta\nu}b^{\alpha}p^{\nu}\gamma^{\mu}\gamma^{\beta}]\nonumber\\
&\,\,&+\Gamma(1-\frac{d}{2})(x(x-1)p^2)^{\frac{d}{2}-1}\epsilon_{\mu\alpha\beta\nu}b^{\alpha}\gamma^{\mu}\gamma^{\beta}\gamma^{\nu}\}\nonumber\\
&=&{e\left(\Gamma\left(d/2\right)\right)^2\over (4\pi)^{{d\over2}}}(-p^2)^{{d\over 2}-2}\left(2ie{\Gamma(2-{d\over 2}) \over \Gamma(d-1)}\slashed{p}-{\Gamma(1-{d\over2})\over \Gamma(d)}  \epsilon_{\alpha\beta\mu\nu}b^{\alpha}((2-d)\slashed{p}\gamma^{\mu}\gamma^{\beta}p^{\nu}-p^2\gamma^{\mu}\gamma^{\beta}\gamma^{\nu})
\right)\nonumber\\
&=&{ie^2\over 8 \pi^2 \epsilon}\slashed{p}-{e\over 48 \pi^2\epsilon}\epsilon_{\alpha\beta\mu\nu}b^{\alpha}(2 p^{\nu}\gamma^{\mu}\gamma^{\beta}\slashed{p}+p^2\gamma^{\mu}\gamma^{\beta}\gamma^{\nu})+\rm{ finite\,\, part}\nonumber\\
&=&{ie^2\over 8 \pi^2 \epsilon}\slashed{p}+{ie\over 24\pi^{2}\epsilon}(5p^2\slashed{b}\gamma^{5}-2(p\cdot b)\slashed{p}\gamma^{5})+\rm{ finite\,\, part}.\label{fermion}
\end{eqnarray}
In the last step, we have used the identity
\begin{equation}
\gamma^{\mu}\gamma^{\nu}\gamma^{\lambda}=\eta^{\mu\nu}\gamma^{\lambda}+\eta^{\nu\lambda}\gamma^{\mu}-\eta^{\mu\lambda}\gamma^{\nu}-i\epsilon^{\sigma\mu\nu\lambda}\gamma_{\sigma}\gamma^{5}. \label{gamma5}
\end{equation}
Note that Eq.~(\ref{gamma5}) can be applied since the dimension dependence of $\gamma_5$ will not affect the results of simple poles in $\epsilon$ and we are only interested in the divergent terms of one-loop corrections in this paper.

The first term in~(\ref{fermion}) is the usual QED correction. The second and third divergent terms indicate that new counterterms of the form $\bar{\Psi} \slashed{b}\gamma^{5}\partial^{2} \Psi$ and $ \bar{\Psi} b\cdot \partial \slashed{\partial}\gamma^{5} \Psi$ are needed. These two terms are not gauge invariant. Later, we will show that by combining all new counterterms, we can rewrite the set of counterterms in terms of gauge invariant operators.  

\subsection{Three-point fermion-photon vertex}
Now we turn our attention to the vertex corrections. The calculation follows the same steps as for the self-energy diagrams. The one-loop contribution to the three-point vertex (Fig. 1c), computed in Feynman gauge, is
\begin{equation}
\Gamma^{\rho}(p',p)=\int {d^d k \over (2\pi)^{d}} {-i g_{\mu\nu} \over k^2} V^{\mu}(k){i\over \slashed{p}'-\slashed{k} } V^{\rho}(q) {i\over \slashed{p}-\slashed{k}} V^{\nu}(-k),
\end{equation}
where $p'^{\mu}=p^{\mu}+k^{\mu}$.

After combining denominators by introducing Feynman parameters and shifting to a new loop momentum variable $l$,  we have
\begin{eqnarray}
\Gamma^{\rho}(p',p)&=&-2e^{2}\int_{0}^{1}\textrm{d}x\textrm{d}y\textrm{d}z\delta(x+y+z-1)\int\frac{\textrm{d}^{d}l}{(2\pi)^{d}}\frac{1}{(l^{2}+y(1-y)p'^{2}+z(1-z)p^{2}-2yzp'\cdot{p})^{3}}\{\nonumber\\
&\,\,&+e \gamma^{\mu}(\slashed{p}'-\slashed{l}-y\slashed{p}'-z\slashed{p})\gamma^{\rho}(\slashed{p}-\slashed{l}-y\slashed{p}'-z\slashed{p})\gamma_{\mu}\nonumber\\
&\,\,&-i(\slashed{l}+y\slashed{p}'+z\slashed{p})(\slashed{p}'-\slashed{l}-y\slashed{p}'-z\slashed{p})\gamma^{\rho}(\slashed{p}-\slashed{l}-y\slashed{p}'-z\slashed{p})\slashed{b}'\nonumber\\
&\,\,&+i\gamma^{\mu}(\slashed{p}'-\slashed{l}-y\slashed{p}'-z\slashed{p})[(b'\cdot{p'}-b'\cdot{p})\gamma^{\rho}-b'^{\rho}(\slashed{p}'-\slashed{p})](\slashed{p}-\slashed{l}-y\slashed{p}'-z\slashed{p})\gamma_{\mu}\nonumber\\
&\,\,&+i\slashed{b}'(\slashed{p}'-\slashed{l}-y\slashed{p}'-z\slashed{p})\gamma^{\rho}(\slashed{p}-\slashed{l}-y\slashed{p}'-z\slashed{p})(\slashed{l}+y\slashed{p}'+z\slashed{p}) \nonumber\\
&\,\,&+2i\epsilon_{\mu \alpha\beta\nu}b^{\alpha}(l+yp'+{z}p)^{\nu}\gamma^{\mu}(\slashed{p}'-\slashed{l}-y\slashed{p}'-z\slashed{p})\gamma^{\rho}(\slashed{p}--\slashed{l}-y\slashed{p}'-z\slashed{p})\gamma^{\beta}\nonumber\\
&\,\,&-i\epsilon^{\rho}_{\,\alpha\beta\nu}b^{\alpha}q^{\nu}\gamma^{\mu}(\slashed{p}'-\slashed{l}-y\slashed{p}'-z\slashed{p})\gamma^{\beta}(\slashed{p}-\slashed{l}-y\slashed{p}'-z\slashed{p})\gamma_{\mu}\}.
\end{eqnarray}
Then, a direct evaluation for the divergent contribution yields
\begin{eqnarray}
\Gamma^{\rho}(p',p)&=&\frac{-ie^{2}}{(4\pi)^{2}\epsilon}\{2e\gamma^{\rho}-2i\epsilon_{\mu\alpha\beta\nu}b^{\alpha}(\gamma^{\mu}\gamma^{\nu}\gamma^{\rho}(\frac{1}{3}\slashed{p}-\frac{1}{6}\slashed{p}')\gamma^{\beta}+\gamma^{\mu}(\frac{1}{3}\slashed{p}'-\frac{1}{6}\slashed{p})\gamma^{\rho}\gamma^{\nu}\gamma^{\beta}\nonumber\\
&\,\,&+(\frac{1}{3}p'+\frac{1}{3}p)^{\nu}\gamma^{\mu}\gamma^{\rho}\gamma^{\beta})
+2i\epsilon^{\rho}_{\,\mu\alpha\beta}b^{\alpha}(p'-p)^{\beta}\gamma^{\mu}\}+\rm{ finite\,\, part}\nonumber\\
&=&\frac{-ie^{3}}{8\pi^{2}\epsilon}\gamma^{\rho}-{ie^2\over 24\pi^2 \epsilon}(5\slashed{b}\gamma^5 (p'+p)^{\rho} - b\cdot(p'+p)\gamma^{\rho}\gamma^5-b^{\rho}(\slashed{p}'+\slashed{p})\gamma^{5} )+\rm{ finite\,\, part}. \label{vertex1}
\end{eqnarray}
Again, the identity~(\ref{gamma5}) has been used in obtaining this result.

The first term in~(\ref{vertex1}) is the usual divergent vertex correction in QED. The other divergent terms in~(\ref{vertex1}) reveal that the LV operators $\bar{\Psi} \gamma^{\mu} b'^{\,\nu} F_{\mu\nu}\Psi$ and $\bar{\Psi}\gamma^{\mu}b^{\nu} \tilde{F}_{\mu\nu} \Psi$ receive no divergent radiative corrections at one-loop order. Instead, LV counterterms of the form $\bar{\Psi}\{ A^{\mu}, \partial_{\mu} \} \slashed{b}\gamma^5 \Psi$
and  $\bar{\Psi}b^{(\mu}\gamma^{\nu)} \{ \partial_{\mu}, A_{\nu}\} \gamma^{5} \Psi$ are required to absorb the new divergences.

\subsection{Four-point fermion-photon vertex}
The radiative corrections to the four-point fermion-photon amplitude in usual QED is finite. However, in the modified LV model~(\ref{L}), this is no longer true. At one-loop order, the four-point fermion-photon vertex receives a correction from the diagram Fig. 1d. In Feynman gauge, the diagram reads
\begin{equation}
\Gamma^{\mu\nu}(p,q_{1},q_{2})=\int {d^d k \over (2\pi)^{d}} {-i g_{\rho\sigma} \over k^2} V^{\rho}(k){i\over \slashed{p}'-\slashed{k} } V^{\mu}(q_{2}) {i\over \slashed{p}-\slashed{k}+\slashed{q}_{1} }V^{\nu}(q_{1}){i\over \slashed{p}-\slashed{k} }V^{\sigma}(-k),
\end{equation}
where $p'^{\mu}=p^{\mu}+q^{\mu}_{1}+q^{\mu}_{2}$.  

Note that due to the symmetry of external photon legs, one only needs to consider the part of this diagram that is symmetric under $\mu\leftrightarrow \nu$. Defining a shifted momentum $l\equiv k-(x+y)p'-(y+z)p$, one can show that the divergent contribution comes from the following integrals:
\begin{eqnarray}
-6i  e^{3}  \int\frac{\textrm{d}^{d}l}{(2\pi)^{d}}\int_{0}^{1}\textrm{d}x\textrm{d}y\textrm{d}z\textrm{d}w{\delta(x+y+z+w-1)\over(l^{2}-\Delta)^{4}} \epsilon_{\alpha\beta\rho\sigma}b^{\alpha}l^{\sigma}(\gamma^{\beta}\slashed{l}\gamma^{\mu}\slashed{l}\gamma^{\nu}\slashed{l}\gamma^{\rho}-\gamma^{\rho}\slashed{l}\gamma^{\mu}\slashed{l}\gamma^{\nu}\slashed{l}\gamma^{\beta}),
\end{eqnarray}
where $\Delta=(({x}+{y})p'+(y+z)p)^{2}-{x}{p'}^{2}-y(p+q_{1})^{2}-{z}p^{2}$.
 
After a straightforward evaluation of the integrals, we arrive at
\begin{eqnarray}
\Gamma^{(\mu\nu)}(p,q_{1},q_{2})= {ie^3\over 24\pi^{2}\epsilon}(5g^{\mu\nu}\slashed{b}\gamma^{5}-b^{(\mu}\gamma^{\nu)} \gamma^{5})+\rm{ finite\,\, part}. \label{vertex2}
\end{eqnarray}
Thus, two counterterms $\bar{\Psi} A^2 \slashed{b}\gamma^{5}\Psi $ and $\bar{\Psi} b\cdot A \slashed{A} \gamma^{5} \Psi$ are required to cancel these divergences.

It is well-known in usual QED that gauge symmetry guarantees the equality of the coefficient of $i\slashed{p}$ in~(\ref{fermion}) with that of $-ie\gamma^{\rho}$ in~(\ref{vertex1}). Here the fact that the coefficient of $p^2\slashed{b}\gamma^{5}$ in~(\ref{fermion}), that of $-e\slashed{b}\gamma^{5}(p'+p)^{\rho}$ in~(\ref{vertex1}), and that of $e^2g^{\mu\nu}\slashed{b}\gamma^{5} $ in~(\ref{vertex2}) are equal and the coefficient of $-2b\cdot p \slashed{p}\gamma^{5}$ in~(\ref{fermion}), that of $e(b\cdot(p'+p)\gamma^{\rho}\gamma^{5}+b^{\rho}(\slashed{p}'+\slashed{p})\gamma^{5})$ in~(\ref{vertex1}), and that of $-e^2b^{(\mu}\gamma^{\nu)}\gamma^{5} $ in~(\ref{vertex2}) are equal is due to the same reason. Therefore, it is easy to show that all new counterterms needed to absorb these divergences can be combined into two gauge invariant operators $\bar{\Psi} D^2 \slashed{b}\gamma^{5}\Psi$ and $\bar{\Psi}b^{\mu} D_{(\mu} D_{\nu)} \gamma^{\nu} \gamma^{5} \Psi$, as it should be. 

In summary, starting with massless QED modified by two non-minimal LV interactions $\bar{\Psi} \gamma^{\mu} b'^{\nu} F_{\mu\nu}\Psi$ and $\bar{\Psi}\gamma^{\mu}b^{\nu} \tilde{F}_{\mu\nu} \Psi$, we have computed all UV divergent one-loop corrections and found out that three additional higher-derivative LV operators $b^{\alpha}F^{\mu\nu} \partial_{\mu} \tilde{F}_{\alpha\nu}$, $\bar{\Psi} D^2 \slashed{b}\gamma^{5}\Psi$ and $\bar{\Psi}b^{\mu} D_{(\mu} D_{\nu)} \gamma^{\nu} \gamma^{5} \Psi$ should be included in the input Lagrangian in order to render quantum corrections finite and keep the predictiveness of the theory at one-loop order.  

\section{$b^{\mu}=0$ case}
From the results of straightforward calculations in the previous section, an interesting consequence is that other than the usual QED divergences, all new divergent corrections are induced by the LV operator $\bar{\Psi}\gamma^{\mu}b^{\nu} \tilde{F}_{\mu\nu} \Psi$. The other operator $\bar{\Psi} \gamma^{\mu} b'^{\nu} F_{\mu\nu}\Psi$ does not contribute to one-loop divergences.  The reason for this is that among all the gauge invariant, $CPT$-violating and $C$-preserving operators that are linear in a fixed vector background, operator $ \bar{\Psi} \gamma^{\mu} b'^{\nu} F_{\mu\nu}\Psi$ is unique in the sense that none of the other operators has the same $P$ and $T$ transformation properties as $ \bar{\Psi} \gamma^{\mu} b'^{\nu} F_{\mu\nu}\Psi$ has. More explicitly, $b'^{0}$ term preserves $P$ parity (and thus violates $T$ parity) and $b'^{i} $ terms preserve $T$ parity (and thus violate $P$ parity). It follows that $\bar{\Psi} \gamma^{\mu} b'^{\nu} F_{\mu\nu}\Psi$ cannot mix with other dimension-five operators by quantum corrections. Then, in the special case where $b^{\mu}=0$ in~(\ref{L}), the results of our analysis in the previous section show that all divergent corrections are the usual QED ones, which can be removed by the renormalization constants and interaction parameters in the original Lagrangian. Hence the theory governed by the Lagrangian
\begin{equation}
\mathcal{L}=-\frac{1}{4}F_{\mu\nu}F^{\mu\nu}+\bar{\Psi}(i\Dslash-\gamma^{\mu}b'^{\nu}F_{\mu\nu})\Psi,  \label{L2} 
\end{equation}
although containg a dimension-five interaction, is one-loop renormalizable.

In this circumstance, given the results of the usual QED one-loop divergences, it is straightforward to determine the renormalization constants $Z_{A,\Psi,e,b'}$, which relate the bare fields, the bare coupling constant, and the bare LV coefficients to the renormalized ones by
\begin{eqnarray}
\Psi_{B}&=&\sqrt{Z_{\Psi} }\Psi,\, \,\,A_{B}^{\mu}=\sqrt{Z_{A}}A^{\mu}, \,\,\,\nonumber \\
e_{B}&=&Z_{e} e,\,\,\, b'^{\,\alpha}_{B}=(Z_{b'})^{\alpha}_{\,\,\,\mu} b'^{\,\mu}.
\end{eqnarray}
The results are:
\begin{eqnarray}
Z_{\Psi}&=&1-{e^2\over 8\pi^2\epsilon},\,\,\, Z_{A}=1-{e^2 \over 6\pi^2\epsilon}, \nonumber\\
 Z_{e}&=& 1+{e^2\over12\pi^2\epsilon},\,\,\, (Z_{b'})^{\alpha}_{\,\,\,\mu}b'^{\,\mu}=b'^{\,\alpha} + {5e^2\over 24\pi^2\epsilon}b'^{\,\alpha}.
\end{eqnarray}

From these renormalization constants, the beta function $\beta_{b'}$ governing the one-loop running of the LV coefficients $b'^{\alpha}$ is found to be
\begin{equation}
(\beta_{b'})^{\alpha} ={5\over 24}{e^2\over\pi^2}b'^{\alpha}.
\end{equation} 
Solving the renormalization group equation, the one-loop running of the LV coefficients $b'^{\alpha}$ is given by 
\begin{equation}
b'^{\alpha}(\mu)=\left(1-{e^2(\mu_{0}) \over 6\pi^2 }\rm{ln}{\mu\over\mu_{0}}\right)^{-{5\over4}} b'^{\alpha}(\mu_{0}). \label{beta}
\end{equation}
This result indicates that the LV coupling $b'^{\alpha}(\mu)$ becomes weaker at low energies. Notice that this running is slow despite the fact that the mass dimension of $b'^{\alpha}$ is negative. In~\cite{loop}, based on the running behaviors of the coefficients associated with LV operators of mass dimension four or less, it is conjectured that there should be a rapid running for LV coefficients with negative mass dimension. However, this is not the case for the theory~(\ref{L2}). In fact, Eq.~(\ref{beta}) tells us that 
\begin{equation}
b'^{\alpha}(M_{Pl})\simeq 1.08\,\, b'^{\alpha}(M_{W}),
\end{equation}
where $M_{Pl}$ and $M_{W}$ are, respectively, the Planck and electroweak scales. This modest running behavior is due to the fact that setting $b'^{\alpha}=0$ enhances the spacetime symmetry group of the specific model~(\ref{L2}), which admits the background vector $b'^{\alpha}$ as an invariant tensor, from $SO(3)$, $SO(2,1)$ or SIM(2) (depending on if $b'^{\alpha}$ is timelike, spacelike, or lightlike\footnote{We thank anonymous referee for the comment on the lightlike case. For more details, see~\cite{Hariton}}, respectively) to the conformal group $SO(4,2)$.

\section{Conclusion}
In this paper, we have computed all UV divergent one-particle-irreducible Feynman diagrams in the LV theory~(\ref{L}) at one-loop order. The divergent corrections to the photon self-energy are given in Eq.~(\ref{photon}), and those to the fermion self-energy are given in Eq.~(\ref{fermion}). The divergent corrections to the three-point and four-point fermion-photon vertices are given in Eq.~(\ref{vertex1}) and Eq.~(\ref{vertex2}), respectively. Our results indicate that other than the usual QED divergences, all new divergent corrections are due to the LV operator $\bar{\Psi}\gamma^{\mu}b^{\nu} \tilde{F}_{\mu\nu} \Psi$. Three additional higher-derivative LV operators $b^{\alpha}F^{\mu\nu} \partial_{\mu} \tilde{F}_{\alpha\nu}$, $\bar{\Psi} D^2 \slashed{b}\gamma^{5}\Psi$ and $\bar{\Psi}b^{\mu} D_{(\mu} D_{\nu)} \gamma^{\nu} \gamma^{5} \Psi$ are found to be required in order to keep the predictiveness of the theory at one-loop order.

We have also shown the one-loop renormalizability of massless QED modified by the operator $\bar{\Psi}\gamma^{\mu}b'^{\nu} F_{\mu\nu} \Psi$, despite the negative mass dimension of the vector background $b'^{\alpha}$. In this circumstance the one-loop beta function for the LV coefficients $b'^{\alpha}$ is determined and solved for the running coefficients. We argue that the slow running of $b'^{\alpha}(\mu)$ between the electroweak and Planck scales described by Eq.~({\ref{beta}) can be understood on the grounds of symmetry. We hope to probe the possible phenomenological applications of this model in the future. 

\begin{acknowledgments}
This research was supported in part by the National Nature Science Foundation of China under Grant No. 10805024 and 555 talent project of Jiangxi Province.
\end{acknowledgments}


\end{document}